\documentclass[12pt,preprint]{aastex}
\usepackage{graphics}
\usepackage{epsfig}
\usepackage{amsmath}
\usepackage{natbib}
\received{July 22 2011}
\revised{August 27 2011}
\accepted{September 16 2011}
\shorttitle{B4: A Pulsating sdB Star in a Binary in NGC~6791}
\shortauthors{Pablo et al.}

\begin{document}
\title{Exploring B4: A Pulsating sdB star, in a Binary, in the Open Cluster NGC~6791}
\author{Herbert Pablo\altaffilmark{1}, Steven D. Kawaler\altaffilmark{1}}
\and
\author{Elizabeth M. Green\altaffilmark{2}}

\altaffiltext{1}{Department of Physics and Astronomy, Iowa State University, Ames, IA  50011, USA}
\altaffiltext{2}{Steward Observatory, University of Arizona, 933 N. Cherry Ave., Tucson, AZ 85721, USA}

\begin{abstract}
We report on {\sl Kepler} photometry of the hot sdB star B4 in the open cluster NGC~6791. We  confirm that B4 is a reflection effect binary with an sdB component and a low-mass main sequence companion with a circular 0.3985~d orbit. The sdB star is a $g-$mode pulsator (a V1093~Her star) with periods ranging from 2384~s to 7643~s. Several of the pulsation modes show symmetric splitting by 0.62$\mu$Hz. Attributing this to rotational splitting, we conclude that  the sdB component has a rotation period of approximately 9.63~d, indicating that tidal synchronization has not been achieved in this system.  Comparison with theoretical synchronization time provides a discriminant between various theoretical models.
\end{abstract}
 

\maketitle

\section{Introduction}
Subdwarf B (sdB) stars are evolved low-mass stars that have helium cores surrounded by a thin hydrogen envelope \citep{saf94}. Their effective temperatures range from 22000 to 40 000 K; typical masses are approximately M$\approx$0.47 $M_{\odot}$ \citep{heb84,heb09}. Since these stars have survived the core helium flash, they provide an opportunity to study a rapid phase of stellar evolution \citep{kawhelas10}, and perhaps a direct probe of the post-flash helium core.

Many sdB stars are non-radial multiperiodic pulsators, which can be used in asteroseismic analysis \citep[and references therein]{charp08,royHelas10}. This analysis can allow us to determine the mass, internal rotation, compositional stratification and other interior properties. There are two main classes of sdB pulsators. 
The first class to be discovered were the shorter period V361~Hya stars \citep{kil97} which are primarily $p-$mode pulsators with  periods typically between 2-4 minutes. 
The V~1093~Her stars are $g$-mode pulsators with periods ranging from  0.75 to 2 h with typical amplitudes of less than 0.1 percent \citep{betsy03, royHelas10}. 
Generally, the pulsation amplitude is higher in the V361~Hya stars (about 1 percent) than in the V1093~Her stars \citep{kil07,reed07}. 

The formation of stars with such thin surface hydrogen layers (less than 0.1\% of the stellar mass) is still not completely understood. There are several proposed formation channels. One channel that has significant observational support involves mass transfer to a companion and ejection of a common envelope \citep{han02,han03}. Observationally, many of the known sdB stars are indeed in close binary systems, with orbital periods on the order of hours \citep{max01,moru03,napi04,heb09}. 

Given the short orbital periods, these stars are generally thought to rotate synchronously as the result of tidal effects. However, two theoretical treatments of tidal synchronization provide a range of estimates for the time scale for synchronization \citep{tas87, Zahn75}; for sdB binaries, they can differ by orders of magnitude.  B4 provides a potential test of these scenarios.
At these rotation velocities, it is not possible to measure rotational broadening in the H or He lines.  The narrow metal lines can show broadening, but the lines are very weak, necessitating high resolution and high S/N spectroscopy requiring large telescopes.  Therefore, spectroscopic verification of tidal synchronization is difficult, though Geier et al. have addressed this issue with the sdB star PG~0101+039 \citep{geier08}.  However, asteroseismology provides a possible way to test for synchronization by measuring rotational splitting of oscillation modes.  In the few cases where this has succeeded, tidal synchronization seems to be verified: \citet{vang08} report that the sdB star Feige 48 appears to be in synchronous rotation.  It is in a binary with a white dwarf companion, with an orbital and rotation period of 9~h.  Another sdB that shows evidence for rotational splitting at the orbital frequency is PG~1336-018 \citep{charp1336}, with an orbital period of 2.42~h and a low-mass main sequence companion.

These successes have been limited by the difficulty faced by ground-based photometry in resolving the pulsation spectra. {\it Kepler} provides long-term nearly continuous high-precision photometry, eliminating aliasing problems associated with ground--based data. Early results from {\it Kepler} observations of sdB stars  
\citep[for example]{roykepI,kaw10,roykepVI} show that {\it Kepler} provides exquisite time series photometry of these stars. For single $g$-mode pulsators, the {\it Kepler} data has already allowed detailed seismic modeling of two sdB stars \citep{vang10,charp11}.  These investigations have determined that the mass of the sdB $g$-mode pulsators is close to what is expected based on standard stellar evolution models.  They also place tight constraints on the hydrogen layer thickness for these stars, and indicate that the convective core may be significantly larger than current evolutionary models suggest.

A particularly interesting star that could shed light on the origins of sdB stars is the hot subdwarf B4 in the old open cluster NGC~6791  \citep{kal92}.  The broadband colors suggested that it was indeed hot enough to be an sdB star.  Spectroscopy by \citet{saf94} confirmed it was an sdB star, and that it was likely a member of NGC~6791 based on its spectroscopic distance.    B4 was identified as a binary through a significant brightness modulation  \citep{moc03,demar07}. However, since no eclipses were seen and the data points were sparse it was not possible to tell whether this was an ellipsoidal or a reflection effect variable.

The temperature determination by \citet{saf94} placed B4 within the range of the V~-1093 Her stars, but it had not been observed with a high enough time resolution to detect pulsations. Photometric variations from binary effects and constraints given its membership in NGC~6791, would make it a uniquely valuable asteroseismic target: a nonradially pulsating sdB star, in a close binary, in a cluster. Its presence within a cluster provides stringent constraints on its age (and metallicity) for comparison with models of sdB formation.  For example, since the lifetime of an sdB star is short ($\sim 10^8$ yr) compared to the cluster age ($\sim 8 \times 10^9$ yr), the mass of the sdB progenitor must be close to the turnoff mass of 1.1 - 1.2 $M_{\odot}$. With this potential in mind, B4 was observed as part of the {\it Kepler} Guest Observer program during Cycle 2.

In this paper we report the discovery of multiperiodic $g-$mode pulsation in B4. We confirm that it shows a longer period variation associated with a reflection of light from a companion with a much lower temperature. This variation signals an orbital period for the binary system of 0,3985~d.  
Our analysis of the pulsations reveals that the sdB star is {\em not} in synchronous rotation.  We also estimate the synchronization time scale using two prescriptions for tidal spin-up, and show that the results favor that of \citet{Zahn75} but are dependent on the details of the structure of the sdB star.

\section{Observations}

While the primary mission of the {\sl Kepler} spacecraft is the search for extrasolar planetary transits \citep{koch10, borucki10}, it is also very well suited for asteroseismic observations \citep{gillast}.  Observations of B4 were obtained in late 2010 and early 2011  by the {\it Kepler} spacecraft during Cycle 2 of the the GO program. In this paper, we analyze the first 6 months  of data (Q6 and Q7) on this star.  The data were taken at the short cadence (SC) mode, with individual integrations of 58.85 s.  The data acquisition and pipeline reductions are as described in \citet{gillsc} and \citet{jonjendata}.  The data coverage is continuous except for short, monthly gaps  for data retrieval and occasional brief safe mode events. The data pipeline produces both a raw and ``corrected`` flux value for each integration. Since the corrected value accounts for estimated background contamination (which may change with follow-up photometry in the future), we use the raw flux. For analysis of the photometric variations, we quote fractional variation from the mean flux.

We removed outliers beyond 4 times the RMS deviation from the local mean (determined via a boxcar filter with a width larger than any expected pulsation but smaller than the timescale for binary variation).  This filtering removed 225 points from the original data. B4 has a V magnitude of 17.87 \citep{demar07} and a  \textit{Kepler} magnitude (K$_{\rm p}$) of 18.27. The noise level is approximately $3.4\times10^{-2}$ per integration; in 6 months of data, this reduces to a noise level of $6.7\times 10^{-5}$.
 
\section{Analysis}

\subsection{Binary Variation}

This star is quite faint even for {\it Kepler}, so to establish a binary ephemeris  we use phased  data to get a better sense of the overall light curve shape. 
We found each point of maximum, rather than keying on another phase, because the light curve shows more sharply--peaked maxima and the time of each maximum could be defined more precisely than the flatter minimum. 
We assumed that the period did not change on time scales smaller than days. This allowed us to add several days of data together with a block-phasing procedure. We achieved the best results by averaging over 10 orbital cycles. Using this procedure we found the period to be 0.3984944(35) d and T0 to be 55372.6002(9) (in BJD - 2400000). 

If this modulation is caused by ellipsoidal variation, then the period of 0.3985 d would represent one half of the orbital period, since in an ellipsoidal variable, there are two minima and two maxima per orbit.  Furthermore, ellipsoidal variables often display minima  of unequal depth 
\citep{hutch74,wil76,boch79}, producing a peak in the Fourier transform at the subharmonic of the highest-amplitude periodicity.

\begin{figure}
\plotone{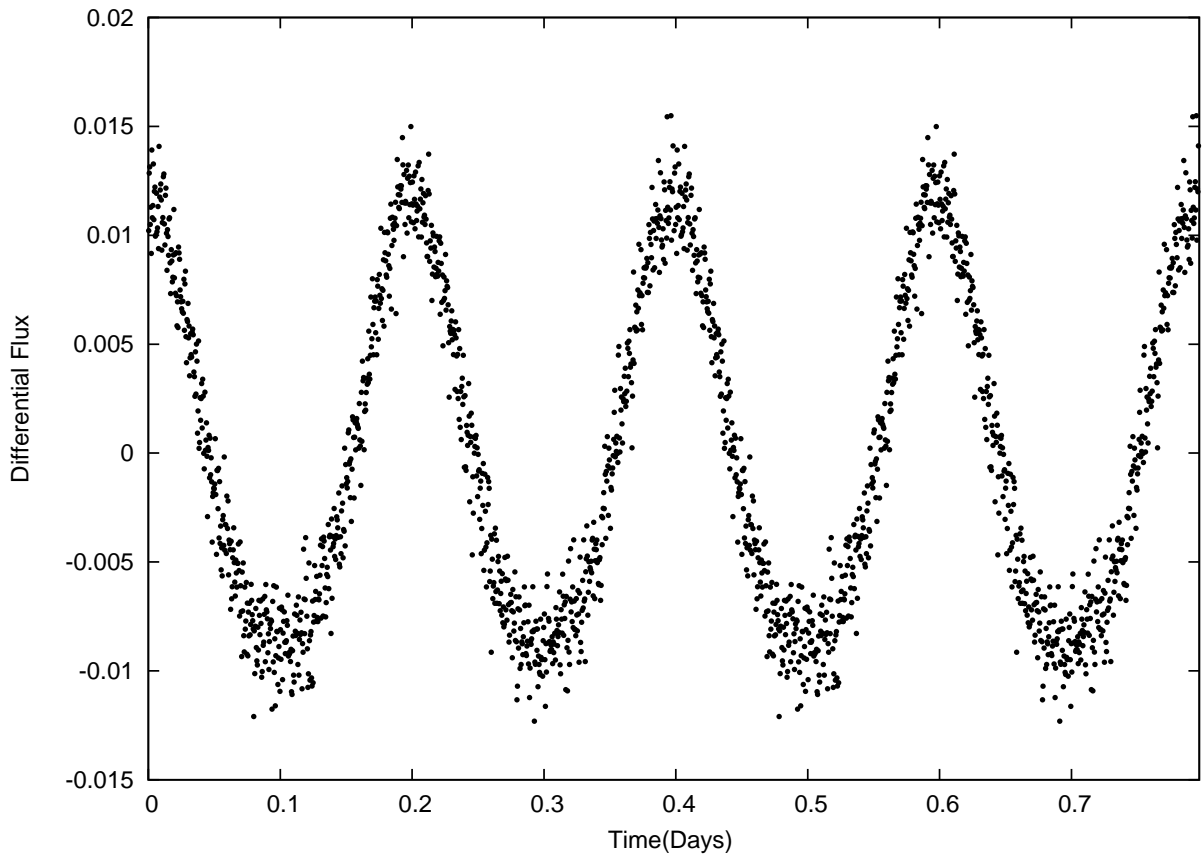}
\caption{Photometric data on B4, phased on twice the suspected orbital period ($2 \times 0.3885$~d).  The equal--depth mimina rule out 
ellipsoidal variation.}
\label{phasedLC}
\end{figure}

Figure \ref{phasedLC} shows the light curve phased at twice the period determined above.This  light curve shows minima of equal depth, and maxima of equal height. There are no eclipses apparent.
In the Fourier transform of the light curve, the largest amplitude peak corresponds to the 0.3985 d period, and the next--highest frequency peak corresponds to the first harmonic of this orbital period. We note that the frequency of the first harmonic of the orbital period (in Table \ref{freqtab}) is slightly more than 1 $\sigma$ away from exactly $2\times f_{\rm orb}$. We do not think that this difference is significant. We see no significant peak at the subharmonic.  

We conclude that the observed variation is caused by the reflection effect, with the light from the sdB star illuminating a cooler and fainter companion.   Thus the period of the binary is 0.3984944(35) d.
A radial velocity curve for this star should therefore show a period of 0.39849 days. In addition, radial velocity observations would constrain the mass ratio of the system.


\subsection{Pulsation}

\begin{figure}
\plotone{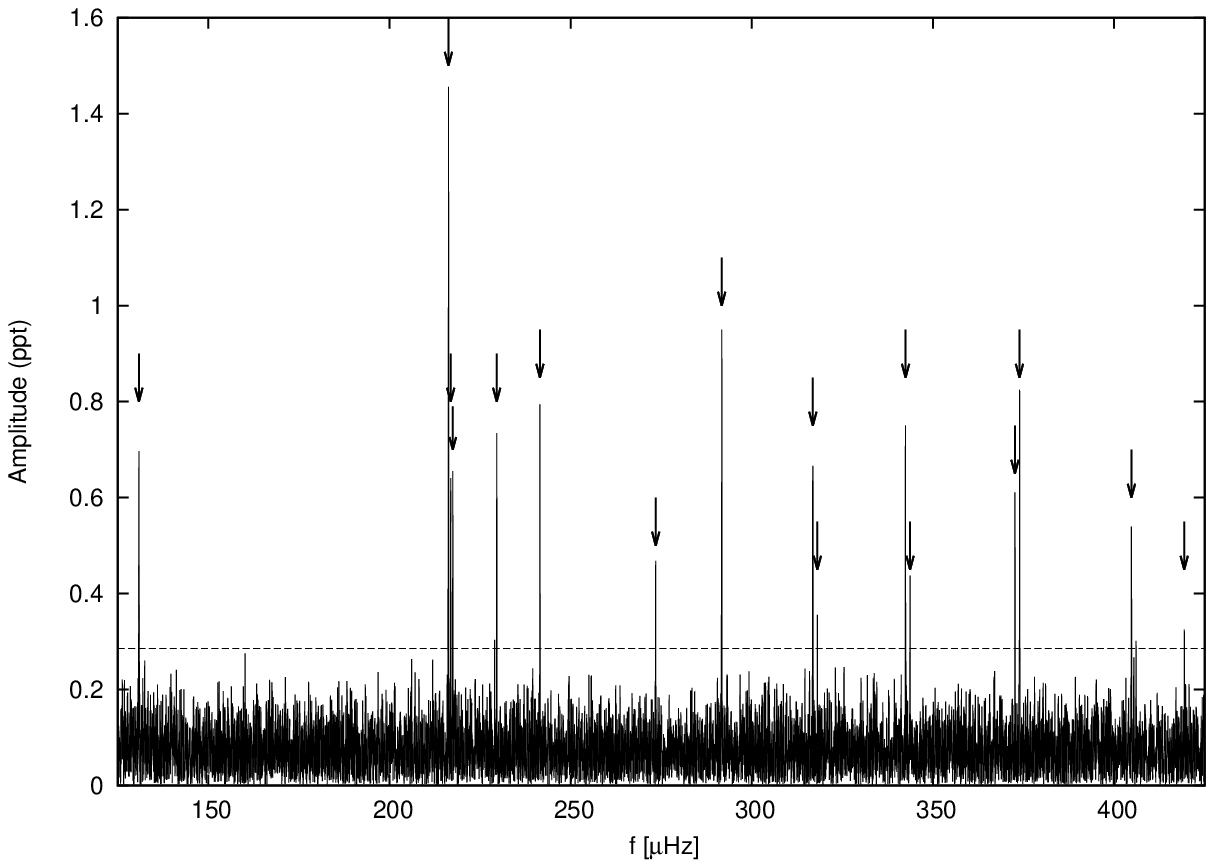}
 \caption{The g-mode region of B4. The arrows show all pre-whitened frequencies The 4 $\sigma$ level above the noise is represented by the dotted line. }
\label{FT-2}
\end{figure}

 The Fourier transform of the $g-$mode region, with the peaks identified, is shown in Figure \ref{FT-2}. B4 shows many periodicities in the frequency range from 120 to 420 $\mu$Hz, which are characteristic of $g-$mode pulsation in V~1093 Her stars.  We followed the ``standard'' procedure of successive removal of periodicities by nonlinear least-squares fitting of sinusoidal signals at the frequency peaks \citep[see][and references therein]{kaw10}.  Each successive prewhitening was performed in the time domain.  We continued this process until no peaks remained at or above 4 times the RMS noise level in the residuals.  This 4-$\sigma$ limit was 0.29 parts per thousand (ppt).   Above this we found 16 unique frequencies which are given in Table \ref{freqtab}. There are 3 more frequencies near the detection threshold which await confirmation with further data, but we include them in Table \ref{freqtab} for reasons noted below.

\begin{deluxetable}{ccccll}
\tabletypesize{\footnotesize}
\tablecaption{Periodicites of B4.  Quoted errors are formal least-squares errors.}
\tablehead{\colhead{ ID} & \colhead{Frequency  [$\mu$Hz]} & \colhead{Period [s]} & \colhead{Amplitude [ppt]} &  \colhead{Orbital splitting} & \colhead{fine structure}}

\startdata
\multicolumn{4}{c}{Binary period and first harmonic} \\
\\
 f$_{\rm orb}$     & 29.04430 $\pm$ 0.00022   & 34430.17 $\pm$ 0.26  & 10.310 $\pm$ 0.064\\                                
2f$_{\rm orb}$     & 58.0901   $\pm$ 0.0014     & 17214.63 $\pm$ 0.42  &  1.579 $\pm$ 0.064 \\ 
                    \\
\multicolumn{6}{c}{Pulsation frequencies} \\
\\
 f1          & 130.8784 $\pm$ 0.0032  & 7640.68   $\pm$ 0.19   &  0.700 $\pm$ 0.064 & &  \\
 f2          & 216.2851 $\pm$ 0.0016  & 4623.527 $\pm$ 0.034 & 1.427 $\pm$ 0.064 & &  \\
 f3          & 216.8996 $\pm$ 0.0038  & 4610.428 $\pm$ 0.068 & 0.081 $\pm$ 0.064 &  &= f2+0.615 \\
 f4          & 217.4811 $\pm$ 0.0036  & 4598.101 $\pm$ 0.076 & 0.631 $\pm$  0.064  & & = f3+0.582 \\
 f5\tablenotemark{1}  & 229.0105                 & 4366.61            &  0.314              & & \\
 f6          & 229.6014 $\pm$ 0.0031  & 4355.374 $\pm$ 0.058  & 0.740 $\pm$ 0.064 &  &= f5+0.591  \\
 f7          & 241.5107 $\pm$ 0.0028  & 4140.604 $\pm$ 0.048  & 0.800 $\pm$ 0.064 & = f3+24.60  & \\
 f8          & 273.4754 $\pm$ 0.0048  & 3656.637 $\pm$ 0.065 & 0.467 $\pm$ 0.064  &=f7 +2$\times$15.98 & \\
 f9          & 291.7086 $\pm$ 0.0024  & 3428.079 $\pm$ 0.028 & 0.948 $\pm$ 0.064 & & \\
 f10        & 316.8387 $\pm$ 0.0034  & 3156.180 $\pm$ 0.034 & 0.664 $\pm$ 0.064  & & \\
 f11        & 318.1220 $\pm$ 0.0062  & 3143.448 $\pm$ 0.062 & 0.363 $\pm$ 0.064 &  & = f10+1.283 \\
 f12        & 342.4285 $\pm$ 0.0030  & 2920.318 $\pm$ 0.026 & 0.744 $\pm$ 0.064 & = f11 + 24.27 &  \\
 f13        & 343.7004 $\pm$ 0.0052  & 2909.511 $\pm$ 0.044 & 0.432 $\pm$ 0.064  & & = f12+1.272  \\
 f14        & 372.6323 $\pm$ 0.0037  & 2683.611 $\pm$ 0.027 & 0.613 $\pm$ 0.064 & &  \\
 f15        & 373.8966 $\pm$ 0.0027  & 2674.536 $\pm$ 0.020 & 0.828 $\pm$ 0.064  &=f12+2$\times$15.73 & = f14+1.264  \\
 f16        & 404.7753 $\pm$ 0.0042  & 2470.506 $\pm$ 0.026 & 0.538 $\pm$ 0.064 & =f15+2$\times$15.44  \\ 
 f17\tablenotemark{1}    & 405.45        & 2466.4        & 0.27 & & =f16+0.68  \\  
 f18\tablenotemark{1}    & 406.04        & 2462.81       & 0.307 & & =f17+0.61  \\ 
 f19	    & 419.3662 $\pm$ 0.0069  & 2384.551 $\pm$ 0.039 & 0.325 $\pm$ 0.064 \\
\enddata
\tablenotetext{1}{Frequencies near the detection threshold; not included in the fit.}
\label{freqtab}
\end{deluxetable}

\begin{figure}
\plotone{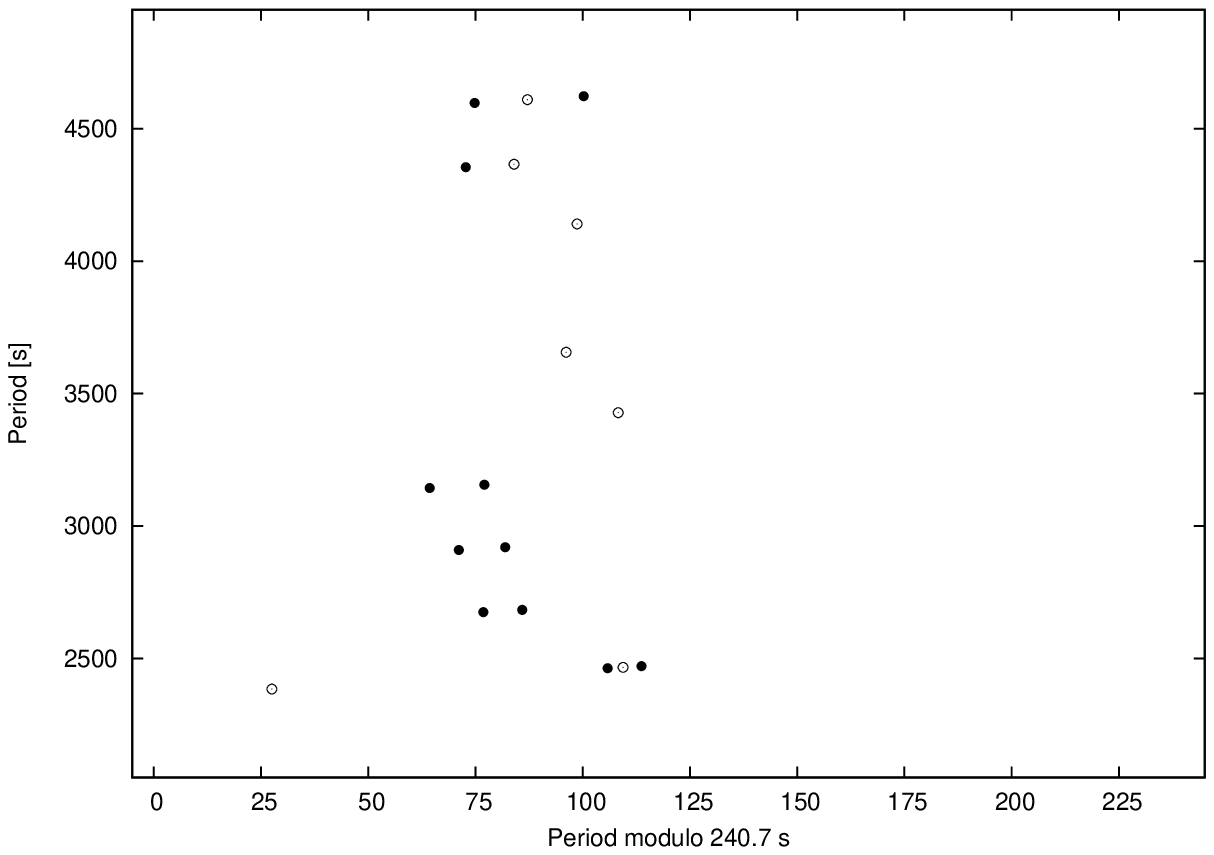}  
\caption{Echelle diagram of the periodicities f2-f18 of B4 with a folding period is 240.7s. Filled circles are suspected $m=\pm 1$ modes, and open circles are $m=0$ modes (or modes for which $m$ cannot be determined).There is a clear ridge around 85 s.}
\label{echelle}
\end{figure}

The $g$-mode period distribution in B4 resembles, quite closely, that seen in non-binary sdB pulsators. In high--overtone $g-$mode pulsators (where $n >> l$), the periods of consecutive overtones should be equally spaced, with a period spacing that scales with $1/\sqrt{l (l+1)}$.  Here, we use the standard labeling of nonradial modes, where $n$, $l$, and $m$ are the radial, angular, and azimuthal quantum numbers.  This behavior is seen in pulsating white dwarf stars \citep[e.g.,][]{winget1159, winget358} and in $g$-mode sdB pulsators \citep{reed11}.  For the sdB stars observed by {\it Kepler} the period spacings range from 231~s to 272~s, with that spacing identified with $l$=1 modes \citep{reed11}.

Figure \ref{echelle} shows that B4 also follows this trend. This echelle diagram plots points associated with each periodicity; the vertical axis is the period and the horizontal axes is the period modulo the average period spacing of 240.7~s.  For equally--spaced modes, the periodicities should line up vertically, with (small) departures to be expected as a result of mode trapping by composition gradients within the star.  The ``best'' period spacing of 240.7~s is in very close accord with $l$=1 pulsations in sdB stars \citep{reed11}. One periodicity, f1 ($P$=7640~s), does not lie near the ridge (it would be at at 226 s on the abscissa of Figure \ref{echelle}). Another lower amplitude mode, f19, also does not fall along the ridge.

As indicated in Figure \ref{echelle} there are multiple periodicities for a given order in the diagram (i.e. 3 successive, nearly horizontal points).  This multiplet structure results from rotational splitting: nonradial modes with the same values of $n$ and $l$ can be split into equally spaced multiplets by rotation, with the frequency splitting proportional to the rotation frequency.   The well--known relationship between frequency splittings and rotation \citep[see, for example][]{ledoux51}, is
\begin{equation}
  f_{n,l,m}=f_{n,l,0}+m\Omega(1-C_{n,l})
\end{equation}  
where $\Omega$ is the (assumed solid-body) rotation frequency, and $C_{n,l}$ is the Ledoux constant,
For B4, we adopt values for $C_{n,l}$ from \citet{kaw10} of 0.48 for $l=1$ and 0.16 for $l=2$ modes.

\begin{figure}
\plotone{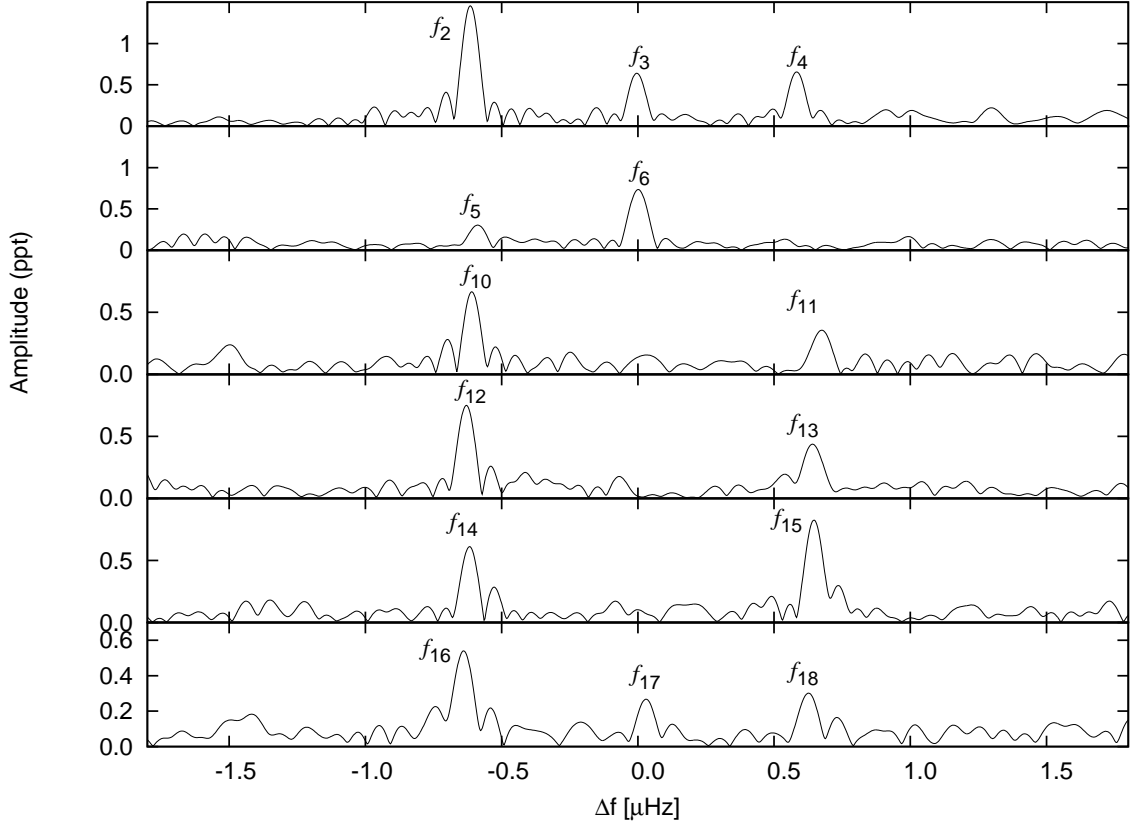}
 \caption{Amplitude spectrum of several $g$-modes in B4, centered on the suspected rotational multiplets. The frequencies of each peak can be found with corresponding labels in Table \ref{freqtab}. There is one triplet (top) with an average spacing of 0.6 $\mu$Hz. Several doublets show twice this splitting. Some of these doublets show signs of a peak halfway in between enhancing the likelihood that 0.6 $\mu$Hz is the rotational splitting.} 
\label{stackedFT}
\end{figure}

As can be seen in Figure \ref{FT-2} and in the table, there is one well-defined triplet in the data with an average spacing of 0.60 $\mu$Hz consisting of f2, f3, and f4. Triplets are expected for rotational splitting of $l$=1 modes.  We also see many doublets with spacing of nearly twice that value: (f10, f11), (f12, f13), and (f14, f15), along with two peaks separated by 0.59 $\mu$Hz (f5, f6).  Taken together, these multiplets, if $l$=1, indicate a rotation frequency of 1.20 $\mu$Hz, (a rotation period of 9.63 d). The structure of the amplitude spectrum surrounding each of these frequencies is shown in Figure \ref{stackedFT}.

B4 is a close binary; if in synchronous rotation, we would expect to see splittings that are approximately 15.4~$\mu$Hz for $l=1$ modes and 24.4~$\mu$Hz for $l=2$ modes.  The measured splittings are much smaller than this orbital frequency, leading to the surprising conclusion that the sdB component is most likely not in synchronous rotation.  Though one might expect the system to be in complete spin--orbit resonance, with the sdB rotating at the orbital frequency, this does not seem to be the case.

The fifth column of Table \ref{freqtab} shows that some of the frequency spacings between periodicities approach the orbital frequency.  This is, we believe, the result of an unfortunate coincidence: the average {\sl period} spacing between $g-$modes of 240 seconds corresponds to a {\sl frequency} spacing ranging from 11.3 $\mu$Hz at the long-period end to 34 $\mu$Hz at the short period side.  Thus some of the apparent frequency spacings that might match those expected for synchronous rotation arise instead as a consequence of equal period spacings for high-overtone $g-$modes.

\subsection{Synchronization time scales}

Recent seismic studies of sdB binary systems with short periods suggest that they are in synchronous rotation;  e.g. \citet{charp1336} and \citet{vang08} looked at systems with orbital periods of 2.42 h and 9.02 h respectively.   \citet{geier08,geier10} claim synchronous rotation in systems with orbital periods up to 14h.  But B4, with an orbital period of  9.56 h does not rotate synchronously.

There are two prescriptions to calculate the time scale for synchronization: \citet{tas87,tas88} and \citet{Zahn75} . \citet{tas87,tas88} argue that large meridional currents driven by tidal effects can drive changes in the rotation rate in nonsynchronous systems. 
For stars with radiative envelopes, \citet{claret95} provides an estimate for the \citet{tas88} synchronization time:
\begin{equation} 
\tau_{\rm syn}=2.13\times 10^4 yr  \left( \frac{1+q}{q}\right) 
                                                              \left(\frac{L}{L_{\odot}}\right)^{-1/4} 
                                                              \left(\frac{M}{M_{\odot}}\right)^{5/4}
                                                              \left( \frac{R}{R_{\odot}}\right)^{-3} 
                                                              \left(\frac{P}{{\rm days}}\right)^{11/4}
\label{tassoul}
 \end{equation}
where $q$ is the mass ratio of the system (assumed to be close to 1), and $P$ is the binary period. For parameters typical of sdB stars ($M=0.48M_{\odot}$, $R \approx 0.2 R_{\odot}$, $L \approx 30 L_{\odot}$), $\tau_{\rm syn} \approx 2\times 10^{5}$ y for $P=0.4$d, assuming a mass ratio of 1. We would thus expect B4 to be synchronous since it has an evolutionary time scale that is a factor of 500 times  longer.  While the \citet{tas88} prescription is not easily extended to low mass ratios, a mass ratio of 0.2 used in Equation \ref{tassoul} yields a synchronization time that is still short compared to the evolutionary time scale.
 
\citet{Zahn75}, explores how tides couple to non-radial oscillations in the star. The oscillations propagate through the convective core and the radiative zone, and provide a torque. 
For stars with radiative envelopes, the resulting synchronization time scale, from \citet{claret97} can be written as
\begin{equation}
\tau_{\rm syn}=3.43 \times 10^6yr \left(\frac{\beta}{0.13}\right)^2 
                                                             \left(\frac{1+q}{q}\right)^2
                                                             \left(\frac{M}{M_{\odot}}\right)^{7/3}
                                                             \left(\frac{R}{R_{\odot}}\right)^{-7}
                                                            \left(\frac{P}{{\rm days}}\right)^{17/3}
                                                            \left(\frac{E_2}{10^{-8}}\right)^{-1}
\label{zahn}
\end{equation}
where $\beta$ the ``radius of gyration'' (the moment of inertia, $I$ scaled by $MR^2$) and $E_2$ is a tidal constant for a given stellar structure.  $E_2$ depends sensitively on stellar structure, and in particular on the fractional size of the convective core.  For large cores, $E_2$ can reach values of $10^{-5}$; for small convective cores, its value approaches zero \citep[see][]{Claret04}. 

Though we do not have values for $E_2$ calculated directly for sdB models,  \citet{Claret04} provides $E_2$ for convective core burning main-sequence models with similar convective core mass fractions to the sdB stars (approximately 0.14).  Though the sdB stars burn helium in the core, the presence or absence of convection is the most important factor for tidal coupling.  \citet{Claret04} main sequence models with 
that size core 
generally have values of $E_2$ between $5\times 10^{-9}$ and $1.3\times 10^{-8}$, somewhat independent of mass.  We choose $10^{-8}$ as a representative value.
Using the same representative values for mass, radius, and orbital period,  Equation \ref{zahn} provides 
$t_{syn} \approx 10^{9}$ yrs for $q=1$. For a lower-mass companion (lower $q$), the time scale is even longer. We note that for Equation \ref{zahn} to reduce to $10^8$ yr for B4, $E_2$ would need to be $\approx 10^{-7}$. For sdB stars with a comparable orbital period but more massive companion (i.e. Feige 48), synchronization should be swifter than for B4.

This leads us to the conclusion that the \citet{Zahn75} mechanism may be slow enough to allow the B4 binary to be out of spin-orbit synchronization.  However, conclusive analysis requires direct calculation of $E_2$ for evolutionary stellar models at the correct  $T_{\rm eff}$ and $\log g$.

\section{Discussion}

The hot blue subdwarf B star, B4, in the old open cluster NGC 6791 is a pulsating member of a
short-period reflection-effect binary with an orbital period of  0.3984944(35) d. Frequency splittings in the $g$-mode pulsation spectrum reveal that the sdB component rotates with a period of 9.63 d, and therefore is not in synchronous rotation.  The nearly--equal period spacings resemble the pattern seen in many other $g$-mode sdB pulsators being observed with {\it Kepler}.  Thus the {\it Kepler} sdB pulsators form a homogenous class, independent of their binarity.

B4's membership in NGC~6791 provides its (overall) age and metallicity; its initial main sequence mass is close to the turnoff mass of 1.15$M_{\odot}$.  Further asteroseismic probing subject to these initial constraints should provide a new opportunity to address the riddle of the formation of sdB stars in general.  Given that B4 is not in synchronous rotation, it will be important to compute the synchronization time scale for sdB stars in close binaries to evaluate the accuracy of current theoretical models of tidal synchronization. A spectroscopic determination of $T_{\rm eff}$ and $\log g$ (as well as the mass ratio of the system) is essential for this. 

We plan continued photometric monitoring of B4 with {\sl Kepler} for as long as possible.  In addition to refining the observed frequencies, increasing S/N could reveal lower-amplitude modes, allowing us to fill out the $l$=1 pulsation spectrum and expose modes with higher values of $l$.  In addition, extended photometry may allow us to measure the evolution of the sdB component through secular period changes as the star continues its nuclear evolution.  If it is a newly--formed sdB star that is experiencing tidal spin-up, we may also be able to measure the increasing rotational frequency through the pulsations.  

\acknowledgments{We thank Andrzej Baran for helpful comments and discussions. Funding for this Discovery mission is provided by NASA's Science
Mission Directorate. The authors gratefully acknowledge the entire {\sl Kepler} team, whose efforts
have made these results possible. 
This material is based upon work supported by the National Aeronautics and Space 
Administration under Grant No. NNX11AC74G issued through the {\sl Kepler} Guest Observer Program - Cycle 2 (09-KEPLER09-0056) to Iowa State University.}

\end{document}